**Andrew Schumann**
Belarusian State University
Andrew.Schumann@gmail.com

# Modal Calculus of Illocutionary Logic

**Abstract:**
The aim of illocutionary logic is to explain how context can affect the meaning of certain special kinds of performative utterances. Recall that performative utterances are understood as follows: a speaker performs the illocutionary act (e.g. act of assertion, of conjecture, of promise) with the illocutionary force (resp. assertion, conjecture, promise) named by an appropriate performative verb in the way of representing himself as performing that act. In the paper I proposed many-valued interpretation of illocutionary forces understood as modal operators. As a result, I built up a non-Archimedean valued logic for formalizing illocutionary acts. A formal many-valued approach to illocutionary logic was offered for the first time.

## 1. A non-Fregean formal analysis of illocutionary acts

Conventional logics including the most non-classical logics like fuzzy logics, paraconsistent logics, etc. satisfy the so-called *Fregean approach* according to that the meaning of a well-formed expression (for instance, the meaning of a propositional formula) should depend on meanings of its components (respectively, on meanings of propositional variables), i.e. the meaning of a composite expression should be a function defined inductively on meanings of atoms. This feature distinguishes the most formal languages from the natural one. In the speech living practice we could exemplify a lot that the meaning of a composite speech act may be not a function defined inductively on meanings of elementary speech acts included in that composite expression. This means that a logic of speech acts cannot satisfy the Fregean approach in general.

Now let us consider the structure of speech acts and their compositions. We are trying to show how compositions in natural language (i.e. in speech acts) differ from compositions defined inductively within the conventional formal logic. Speech acts from the logical point of view are said to be *illocutionary acts*. We know that whenever a speaker utters a sentence in an appropriate context with certain intentions, he performs one or more illocutionary acts. We will denote the simple illocutionary act by $F(\Phi)$.

This denoting means that each simple illocutionary act can be regarded as one consisting of an illocutionary force F and a propositional content Φ. For example, the utterance "I promise you (F) to come (Φ)" has such a structure. Usually, the illocutionary force is expressed by performative sentences which consist of a performative verb used in the first or third person present tense of the indicative mood with an appropriate complement clause. In our case we used the example of illocutionary act with the performative verb "promise." However, the illocutionary force is not totally reduced to an appropriate performative verb. It also indicates moods of performance (e.g., moods of order are distinguished in the following two sentences: "Will you leave the room?", "If only you would leave the room"). Notice that an illocutionary force of performative verb can be expressed nonverbally too, e.g. by means of intonations or gestures. Traditionally, the illocutionary force is classified into five groups that are called illocutionary points: assertives, commissives, directives, declaratives, expressives.[1]

The illocutionary points are used in the setting of simple illocutionary acts. When a simple illocutionary act is successfully and non-defectively performed there will always be effects produced in the hearer, the effect of understanding the utterance and an appropriate perlocutionary effect (for instance, the further effect on the feeling, attitude, and subsequent behavior of the hearer). Thus, an illocutionary act must be both successful and non-defective. Recall that in classical logic a well-formed proposition is evaluated as either true or false and in conventional logics as a degree of truth (the latter could be however very different, e.g. it could run the unit interval [0, 1] as in fuzzy logics, trees of some data as in spatial logics and behavior logics, sets of truth values as in higher-order fuzzy logics and some paraconsistent logics, etc.). But as we see, a non-defective simple illocutionary act is evaluated as *either successful or unsuccessful* in the given context of utterance (notice that within the more detailed consideration in the same way as in non-classical logics, the meaning of non-defective simple illocutionary act could be evaluated as a degree of successfulness).

---

[1] In short, they are distinguished as follows: "One can say how things are (assertives), one can try to get other people to do things (directives), one can commit oneself to doing things (commissives), one can bring about changes in the world through one's utterances (declarations), and one can express one's feelings and attitudes (expressives)" (Searle and Vanderveken 1984: 52).

Regarding compositions in speeches, i.e. appropriate composite illocutionary acts built up from simple ones, we could distinguish two cases: first, the case of compositions that satisfy the Fregean approach and, second, the case of compositions that break the latter. The criterion will be as follows. If a logical superposition of simple illocutionary acts may be also evaluated as either successful or unsuccessful, it is said to be a *complex illocutionary act*. In this case we have no composition defined inductively. If it may be examined as either true or false, then this logical superposition is said to be an *illocutionary sentence*. In this case we can follow the Fregean approach.

Logical connectives ($\vee, \wedge, \Rightarrow, \neg$) which we use in the building of complex acts or sentences we will call illocutionary connectives. They differ from usual ones. As an example, the illocutionary disjunction in the utterance "I order to leave the room or I order to not leave the room" differs from the usual propositional disjunction, because it doesn't express here the law of excluded middle and a hearer can reject it as an unsuccessful illocutionary act.

Sometimes a logical superposition of simple illocutionary acts with propositions also sets a complex illocutionary act. Some examples are as follows: "If it rains, I promise you I'll take my umbrella" (the illocutionary implication of the form $\Psi \Rightarrow F(\Phi)$), "It rains and I assert that I'll take my umbrella" (the illocutionary conjunction of the form $\Psi \wedge F(\Phi)$). But we can consider cases when a logical superposition of simple illocutionary acts with propositions doesn't get a complex illocutionary act. As an example, "If I think so, then really it is so" (the illocutionary implication of the form $F(\Phi) \Rightarrow \Phi$). Indeed, it is a true illocutionary sentence.

Thus, I distinguish illocutionary sentences from illocutionary acts. Just as illocutionary acts express appropriate performances, so illocutionary sentences say about logical properties of illocutionary acts. Existence of illocutionary sentences (i.e. some inductive compositions in composite speeches) allows us to set a special logic that is called *illocutionary logic*. It studies logical and semantic properties of illocutionary acts and illocutionary sentences. Therefore "just as propositional logic studies the properties of all truth functions …, so illocutionary logic studies the properties of illocutionary forces without worrying about the various ways that these are realized in the syntax of English" (Searle and Vanderveken 1984). Illocutionary logic plays an important role in modern analytical philosophy of language and in logical models of speech acts: its aim is to explain how context can affect the meaning of certain special kinds of illocutionary acts. The first formalization of illocutionary logic was created by J.R. Searle and D. Vanderveken (1984). In that work, a semantic-phenomenological approach was proposed and in the framework of this approach all the conditions of success and non-defection of illocutionary acts were precisely investigated.

In this paper I propose a logical-syntactic approach to illocutionary logic, according to that I consider illocutionary forces as modal operators which have a many-valued interpretation. This approach can supplement the approach of J.R. Searle and D. Vanderveken.

## 2. A many-valued illocutionary logic with the only performative verb "think"

In order to show how we could set up compositions in illocutionary logic and combine Fregean and non-Fregean compositions within a logical system in general, we first try to define a propositional logic with the only performative verb "think." Let us consider a propositional language L that is built in the standard way with the additional unary operator F. It is called the illocutionary force of the performative verb "think." We will say that the illocutionary act F($\Phi$) is a performance of a proposition $\Phi$. From the point of view of social constructivism (Berger 1971), the content of social acts and the content of performances of any propositions are not physical facts. Therefore performances cannot be evaluated as either true or false.

The performance of $\Phi$ that we obtain by using the performative verb "think" can be either successful or unsuccessful. It is successful if F($\Phi$) represents a true propositional content of $\Phi$ (i.e., if $\Phi$ is true) and it is not successful if F($\Phi$) represents a false propositional content of $\Phi$ (i.e., if $\Phi$ is false). The success and unsuccess of performance will be denoted by 1/2 and −1/2 respectively.

Further, let us suppose that atomic propositions, i.e. propositional variables, (they belong to the set Var$L$ := {p, $p_1$, $p_2$, …}) can have only one of the following two truth values: 1 ("true") and 0 ("false"). Let our language *L* be associated with the following matrix M = <{1, 1/2, 0,−1/2}, {1}, ¬, F, ⇒, ∨, ∧ >, where

- {1, 1/2, 0,−1/2} is the set of truth values
- {1} is the singleton of designated truth values,
- ¬, F are unary operations for negation and illocutionary force respectively, both are defined as follows:

$$\neg x = \begin{cases} 1 - x \text{ if } x \in \{1, 0\}, \\ -x \text{ if } x \in \{\frac{1}{2}, -\frac{1}{2}\}; \end{cases}$$

$$F(x) = \begin{cases} x - \frac{1}{2} \text{ if } x \in \{1, 0\}, \\ x \text{ if } x \in \{\frac{1}{2}, -\frac{1}{2}\}. \end{cases}$$

- ⇒, ∨, ∧ are binary operations for disjunction, conjunction, and implication respectively, their definitions:

$$x \Rightarrow y = 1 - \sup(x, y) + y = \begin{cases} 1 \text{ if } x \leq y, \\ y \text{ if } x > y \text{ and } x = 1, \\ \frac{1}{2} \text{ if } x > y \text{ and } x = 0 \text{ or } y = 0, \\ 0 \text{ otherwise;} \end{cases}$$

$$x \vee y = \begin{cases} \inf(x, y) \text{ if } x, y \in \{\frac{1}{2}, -\frac{1}{2}\}, \\ \sup(x, y) \text{ otherwise;} \end{cases}$$

$$x \wedge y = \begin{cases} \sup(x, y) \text{ if } x, y \in \{\frac{1}{2}, -\frac{1}{2}\}, \\ \inf(x, y) \text{ otherwise.} \end{cases}$$

Then it could easily be proved that the unary operator F satisfies the following conditions:

$$\forall a \in M. \ a \geq F(a), \tag{1}$$

$$\forall a \in M.\ \neg a \geq \neg F(a), \qquad (2)$$

$$\forall a, b \in M.\ (F(a) \wedge F(b)) \geq F(a \wedge b), \qquad (3)$$

$$\forall a, b \in M.\ (F(a) \vee F(b)) \leq F(a \vee b), \qquad (4)$$

$$\forall a, b \in M.\ (F(a) \Rightarrow F(b)) \geq F(a \Rightarrow b), \qquad (5)$$

$$\forall a \in M.\ F(F(a)) = F(a), \qquad (6)$$

$$\forall a \in M.\ \neg F(a) = F(\neg a). \qquad (7)$$

Let $e$ be a valuation of atomic propositions, i.e. $e$: Var$L \to \{0, 1\}$. We can extend $e$ to the case of the following valuation Ve : L $\to \{1, 0, 1/2, -1/2\}$ by using the operations of M which are assigned to appropriate logical connectives. The valuation Ve is called an *illocutionary valuation*.

Let $\Phi \in L$. The performance of $\Phi$, i.e. $F(\Phi)$, is called an unsuccessful for $e$ if Ve($F(\Phi)$) = $-1/2$, i.e. Ve($\Phi$) $\in \{0, -1/2\}$. The formula $F(\Phi)$ is called a successful performance for $e$ if Ve($F(\Phi)$) = $1/2$, i.e. Ve($\Phi$) $\in \{1, 1/2\}$. Further, the formula $\Phi$ is called a true sentence for $e$ if Ve($\Phi$) = 1 and a false sentence for $e$ if Ve($\Phi$) = 0.

Notice that the element a $\wedge$ $\neg$a is not a minimal in M, because a $\wedge$ $\neg$a $\geq$ F(a $\wedge$ $\neg$a) and (F(a) $\wedge$ $\neg$F(a)) $\geq$ F(a $\wedge$ $\neg$a). Consequently, the minimal element of M (that is called 'illocutionary contradiction' or 'unsuccess of performance') is assigned to a sentence of the form F($\Phi$ $\wedge$ $\neg\Phi$) when somebody thinks a propositional contradiction.

Let us show that the illocutionary valuation defined above satisfies the informal understanding of the concept 'illocutionary act.' First of all, we should notice that the illocutionary force of "think" sets up a warp of the logical space of propositional relations (by analogy with the gravitational force which warps the physical space). For instance, suppose that there are two successful illocutionary acts: "I think that if it is harmful for my health, then I will not do it" and "I think that smoking is harmful for my health." Both don't imply that the illocutionary act "I think that I will not smoke" is successful. However, we could entail according to modus ponens if the illocutionary force of the verb "think" was removed in both expressions. So, we can claim that the illocutionary force may be considered in terms of the warpage of logical space in the same way as the gravitational force has been regarded recently in terms of the warpage of Euclidean space.

This feature is illustrated by inequalities (1) – (7) that could be converted to equalities if we removed the illocutionary force F. Namely, inequality (1) means that the implication $F(\Phi) \Rightarrow \Phi$ is a true sentence of illocutionary logic. For example, "If I think that he is God, then he is God" (but not vice versa) is an example of illocutionary tautology. Formula (2) means that the implication $\neg F(\Phi) \Rightarrow \neg \Phi$ is also a true illocutionary sentence. Indeed, something exists if I think so and something doesn't exist if I don't think so. Inequality (3) means that the implication $F(\Phi \wedge \Psi) \Rightarrow (F(\Phi) \wedge F(\Psi))$ is a tautology of illocutionary logic. For instance, the following illocutionary sentence is true: "If she thinks that the weather is good and the world is fine, then she thinks that the weather is good and she thinks that the world is fine" (but not vice versa). Continuing in the same way, we can see that formulas (4) – (6) satisfy an intuitive meaning, too. Notice that the verb "think" readily differ from others within formula (7). The illocutionary act "I think that it is not white" is equivalent to "I don't think that it is white." In the meantime, an illocutionary act with the negation of an illocutionary force that we obtain using other performative verbs can differ from the same illocutionary act with a positive illocutionary force, but with the negative propositional content. An example of illocutionary negation: "I do not promise to come", an example of an illocutionary act with a negative propositional content: "I promise not to come." As we see, these acts are different.

I have just exemplified how we can combine Fregean and non-Fregean compositions of simple illocutionary acts within a logical system. I have done this by distinguishing composite expressions evaluated as 0 and 1 from composite expressions evaluated as 1/2 and −1/2. The first expressions correspond to the Fregean approach (when the meaning of the whole is reduced to its components), the second ones evidently do not (because they assume to be ordered by the dual relation, where conjunction is understood as supremum and disjunction as infimum; as a result, the meaning of composite expression changes meanings of components:

conjunction of components starts to be interpreted as disjunction on the level of the whole expression and vice versa). Taking into account that we had postulated the existence of the only performative verb, we defined illocutionary logical operations by induction. In this meaning, we built up a many-valued illocutionary logic of Frege's style and did our best to present both Fregean and non-Fregean compositions within conventional logic. However, it was just an illustration in the simplest case of using the only performative verb. This exemplification allowed us to clarify that non-Fregean compositions in speech acts can be regarded logically too, by a kind of dualization, when meanings start to be ordered by the dual relation.

**3. An ordering relation on the set of all illocutionary forces**
Now set up the more complicated problem to construct a many-valued illocutionary logic closed under the class of all performative verbs. We can be sure that in this case our logic will not be of Frege's style. In the beginning, we should define a *partial ordering relation* $\leq$ on an infinite class of all illocutionary forces (expressed by all performative verbs). This allows us to define the basic logical connectives ($\vee, \wedge, \Rightarrow, \neg$) on the set of simple illocutionary acts. Intuitively, the relation $Ve(F_1(\Phi)) \leq Ve(F_2(\Psi))$, where Ve is supposed to be an evaluation function run over well-formed formulas (however, it is not defined yet), holds if and only if an illocutionary act $F_1(\Phi)$ logically entails an illocutionary act $F_2(\Psi)$. In the latter case it is not possible to perform the first act without thereby performing the second (i.e. for all contexts of utterances $i \in I$, if $F_1(\Phi)$ is performed at $i$, then $F_2(\Psi)$ is performed at $i$, too).

A new ordering relation also provides us with a possibility to consider so-called *disjoint* or *opposite illocutionary acts* $F_1(\Phi), F_2(\Phi)$. Both are disjoint/opposite when $F_1(\Phi) \wedge F_2(\Phi)$ is not simultaneously performable, more precisely when the successful performance of $F_1(\Phi)$ is relatively inconsistent (1) with the achievement of the illocutionary point of $F_2$ on $\Phi$, or (2) with the degrees of strength of illocutionary point of $F_2$, or (3) with the satisfaction by $\Phi$ of the propositional content conditions of $F_2$, or (4) with the presupposition of the preparatory conditions of $F_2(\Phi)$, or finally (5) with a commitment to the psychological state of $F_2(\Phi)$.

In the next section, we are trying to examine general properties of opposite illocutionary acts.

## 4. A square of opposition for illocutionary acts

First, assume that two illocutionary forces $F_1$ and $F_2$ are stand in the semantical opposition. This means that they cannot at the same time both be successful, however both may be unsuccessful. In other words, an illocutionary act $F_1(\Phi)$ is semantically equivalent to an illocutionary act $F_2(\neg\Phi)$ (clearly, in this case $F_2(\Phi)$ is semantically equivalent to $F_1(\neg\Phi)$), i.e. $Ve(F_1(\Phi)) = Ve(F_2(\neg\Phi))$ and $Ve(F_1(\neg\Phi)) = Ve(F_2(\Phi))$. For example, the performative verbs "order" and "forbid" have such opposition. The *square of opposition* or *logical square for illocutionary acts* is a natural way of classifying illocutionary forces which are relevant to a given opposition. Starting from two illocutionary forces $F_1$ and $F_2$ that are contraries, the square of opposition entails the existence of two other illocutionary forces, namely $\neg F_1$ and $\neg F_2$. As a result, we obtain the following four distinct kinds of opposition between pairs of illocutionary acts:

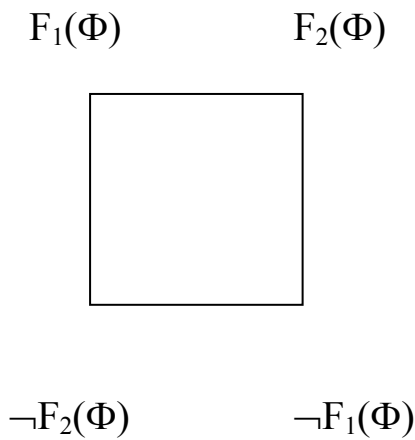

**Figure 1**. *The square of oppositions for illocutionary forces $F_1$, $F_2$. For instance, $F_1$ is "bless" and $F_2$ is "damn." There can be an infinite number of such squares for different illocutionary forces. At least, there exists an appropriate square of opposition for each illocutionary force. In the latter case $F_1$ is "bless to do" and $F_2$ is "bless not to do."*

- Firstly, illocutionary forces $F_1$ and $F_2$ are *contrary*. This means that they cannot be successful together in corresponding illocutionary acts $F_1(\Phi)$ and $F_2(\Phi)$ with the same propositional content, but as we see both may be unsuccessful: "I insist that he takes part" and "I insist that he doesn't take part." A speaker cannot successfully simultaneously say: "I order you to leave the room" and "I forbid you to leave the room."
- Secondly, illocutionary forces $F_1$ and $\neg F_1$ (resp. $F_2$ and $\neg F_2$) are *contradictory*, i.e. the success of one implies the unsuccess of the other, and conversely. For instance, the success of the illocutionary act "I order you to do this" implies the unsuccess of the corresponding act "I request that you do not do it," and conversely.
- Next, illocutionary forces $\neg F1$ and $\neg F2$ are *subcontrary*, i.e. it is impossible for both to be unsuccessful in corresponding illocutionary acts $\neg F_1(\Phi)$ and $\neg F_2(\Phi)$ with the same propositional content, however it is possible both to be successful.
- Lastly, illocutionary forces $F_1$ and $\neg F_2$ (resp. $F_2$ and $\neg F_1$) are said to stand in the *subalternation*, i.e. the success of the first ("the superaltern") implies the success of the second ("the subaltern"), but not conversely. As an example, the success of the illocutionary act "he orders" implies the success of the illocutionary act "he requests", but not vice versa. Consequently, the success of an illocutionary act $F_1(\Phi)$ or $F_2(\Phi)$ implies the success of the corresponding illocutionary act $\neg F_2(\Phi)$ or $\neg F_2(\Phi)$, respectively. Consequently, the unsuccess of an illocutionary act $\neg F_2(\Phi)$ or $\neg F_2(\Phi)$ implies the unsuccess of the corresponding illocutionary act $F_1(\Phi)$ or $F_2(\Phi)$, respectively.

Thus, pairs of illocutionary acts are called contradictories (contradictoriae) when they cannot at the same time both be successful or both be unsuccessful, contraries (contrariae) when both cannot at the same time be successful, subcontraries (subcontrariae) when both cannot at the same time be unsuccessful, and subalternates (subalternae) when the success of the one act implies the success of the other, but not conversely.

There is a very easy way to detect whether the square of opposition for the given illocutionary acts holds. The expected criterion is as follows: if $F(\neg\Phi) \Rightarrow \neg F(\Phi)$ is a true illocutionary sentence, i.e. $Ve(F(\neg\Phi)) \leq Ve(\neg F(\Phi))$, then the square of opposition for illocutionary acts holds. As a corollary, the following expressions are to be regarded as true illocutionary sentences:

$$\neg F(\neg\Phi) \vee \neg F(\Phi) - \text{tertium non datur,} \qquad (8)$$

$$\neg(F(\neg\Phi) \wedge F(\Phi)) - \text{the law of contrary.} \qquad (9)$$

## 5. A non-Archimedean interpretation of illocutionary forces

Now we are trying to develop a formal approach to evaluations of illocutionary forces due to non-Archimedean semantics proposed by Schumann (2008) and Schumann (2009). Suppose B is a complete Boolean algebra with the bootom element 0 and the top element 1 such that the cardinality of its domain |B| is an infinite number. Build up the set $B^B$ of all functions $f : B \rightarrow B$. The set of all complements for finite subsets of B is a filter and it is called a Frechet filter, it is denoted by U. Further, define a new relation $\approx$ on the set $B^B$ by $f \approx g = \{a \in B : f(a) = g(a)\} \in U$. It is easily be proved that the relation $\approx$ is an equivalence. For each $f \in B^B$ let $[f]$ denote the equivalence class of f under $\approx$. The ultrapower $B^B/U$ is then defined to be the set of all equivalence classes $[f]$ as f ranges over $B^B$. This ultrapower is called a *nonstandard* (or *non-Archimedean*) *extension* of B, for more details see Robinson (1966) and Schumann (2008). It is denoted by *B.

There exist two groups of members of *B: (1) functions that are constant, e.g. $f(a) = m \in B$ on the set U, a constant function $[f = m]$ is denoted by *m, (2) functions that aren't constant. The set of all constant functions of *B is called standard set and it is denoted by °B. The members of °B are called standard. It is readily seen that B and °B are isomorphic.

We can extend the usual partial order structure on B to a partial order structure on °B:
- for any members x, y ∈ B we have $x \leq y$ in B iff $*x \leq *y$ in °B,
- each member $*x \in °B$ (which possibly is a bottom element *0 of °B) is greater than any number $[f] \in *B \backslash °B$, i.e. $*x > [f]$ for any $x \in B$, where $[f]$ isn't constant function.

Notice that under these conditions, there exist the top element $*1 \in *B$ such that $1 \in B$, but the element $*0 \in *B$ such that $0 \in B$ is not bottom for $*B$.

The ordering conditions mentioned above have the following informal sense: (1) the sets $°B$ and $B$ have isomorphic order structure; (2) the set $*B$ contains actual infinities that are less than any member of $°B$. These members are called Boolean infinitesimals.

Introduce three operations 'sup', 'inf', '¬' in the partial order structure of $*B$:

$$\inf([f], [g]) = [\inf(f, g)];$$

$$\sup([f], [g]) = [\sup(f, g)];$$

$$\neg[f] = [\neg f].$$

This means that a nonstandard extension $*B$ of a Boolean algebra $B$ preserves the least upper bound 'sup', the greatest lower bound 'inf', and the complement '¬' of $B$.

Consider the member $[h]$ of $*B$ such that $\{a \in B: h(a) = f(\neg a)\} \in U$. Denote $[h]$ by $[f\neg]$. Then we see that $\inf([f], [f\neg]) \geq *0$ and $\sup([f], [f\neg]) \leq *1$. Really, we have three cases.
- *Case 1*. The members $\neg[f]$ and $[f\neg]$ are incompatible. Then $\inf([f], [f\neg]) \geq *0$ and $\sup([f], [f\neg]) \leq *1$,
- *Case 2*. Suppose $\neg[f] \geq [f\neg]$. In this case $\inf([f], [f\neg]) = *0$ And $\sup([f], [f\neg]) \leq *1$.
- *Case 3*. Suppose $\neg[f] \leq [f\neg]$. In this case $\inf([f], [f\neg]) \geq *0$ And $\sup([f], [f\neg]) = *1$.

Now define hyperrational valued matrix logic $M_B$ as the ordered system $<*B, \{*1\}, \neg, \Rightarrow, \vee, \wedge>$, where
- $*B$ is the set of truth values,
- $\{*1\}$ is the set of designated truth values,
- for all $[x] \in *B$, $\neg[x] = *1 - [x]$,
- for all $[x], [y] \in *B$, $[x] \Rightarrow [y] = *1 - \sup([x], [y]) + [y]$,

- for all *x, *y ∈ °B, *x ∨ *y = sup(*x, *y),
- for all *x, *y ∈ °B, x ∧ y = inf(*x, *y).
- for all [x], [y] ∈ *B\°B, [x] ∨ [y] = inf([x], [y]),
- for all [x], [y] ∈ *B\°B, [x] ∧ [y] = sup([x], [y]).

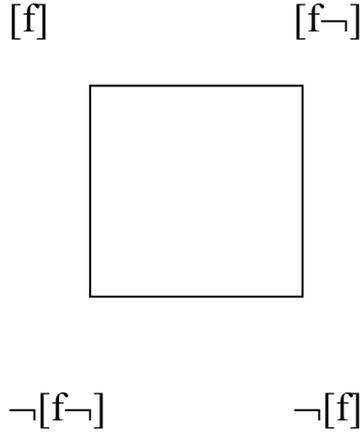

[f]               [f¬]

¬[f¬]            ¬[f]

**Figure 2**. In case [f¬] ≤ ¬[f], *the square of oppositions for any members* [f], [f¬], ¬[f¬], ¬[f] *of* *B *holds true, i.e.* [f], [f¬] *are contrary,* [f], ¬[f] *(resp.* ¬[f¬], ¬[f]*) are contradictory,* ¬[f¬], ¬[f] *are subcontrary,* [f], ¬[f¬] *(resp.* [f¬], ¬[f]*) are said to stand in the subalternation .*

Let us consider a propositional language $L_B$ that is built in the standard way with the additional set of unary operators $F_1$, $F_2$, $F_3$, ... We suppose that for each performative verb, this set contains an appropriate sign. So, this set should be uncountable infinite. We associate the language $L_B$ with the matrix $M_B$ in accordance with the illocutionary evaluation Ve satisfying the following properties:

- Ve(F(Ψ)) = [f] ∈ *B\°B if Ψ does not contain unary operators $F_1$, $F_2$, $F_3$, ...;
- Ve($F_i$($F_j$(Ψ))) = ([f]$_i$ ⇒ Ve($F_j$(Ψ))), where [f]$_i$ ∈ *B\°B and ⇒ is an operation of $M_B$.;
- Ve(Ψ) = *x ∈ °B if Ψ does not contain unary operators $F_1$, $F_2$, $F_3$, ...;
- Ve(Φ ∧ Ψ) = (Ve(Φ) ∧ Ve(Ψ)), where on the right-side ∧ is an operation of $M_B$;

- Ve(Φ ∨ Ψ) = (Ve(Φ) ∨ Ve(Ψ)), where on the right-side ∨ is an operation of $M_B$;
- Ve(Φ ⇒ Ψ) = (Ve(Φ) ⇒ Ve(Ψ)), where on the right-side ⇒ is an operation of $M_B$.

It can easily be proved that the following formulas are illocutionary tautologies according to the matrix $M_B$ in the same measure as they were in the matrix M (see section 2):

$$F(\Phi) \Rightarrow \Phi, \qquad (10)$$

$$\neg F(\Phi) \Rightarrow \neg \Phi, \qquad (11)$$

$$F(\Phi \wedge \Psi) \Rightarrow (F(\Phi) \wedge F(\Psi)), \qquad (12)$$

$$(F(\Phi) \vee F(\Psi)) \Rightarrow F(\Phi \vee \Psi), \qquad (13)$$

$$F(\Phi \Rightarrow \Psi) \Rightarrow (F(\Phi) \Rightarrow F(\Psi)), \qquad (14)$$

Instead of formula (7), the matrix $M_B$ satisfies the relations of square of oppositions for illocutionary acts. Formula (6) does not hold in general, too, because each illocutionary force effects in a unique way. For instance, a cyclic self-denying illocutionary act "I promise I will not keep this promise" F(¬F(F(¬F(…)))) is evaluated as follows: Ve(F(¬F(F(¬F(…))))) = ([f] ⇒ (¬[f] ⇒ ([f] ⇒ (¬[f] ⇒ (…))))). As we see, a cyclic self-denying illocutionary act does not correspond to any member of *B. The matter is that we have there an infinite sequence ([f] ⇒ (¬[f] ⇒ ([f] ⇒ (¬[f] ⇒ (…))))) whose whole truth-value depends on a truth-value of the most right atomic formula, but the sequence is infinite and we have no end-right atomic formula.

The novel logic is not Fregean, though it is evidently formal and furthermore it can be shown that some restrictions of this logic may be complete (Schumann 2008). The point is that we obtain an uncountable infinite set of well-formed formulas for that there is no induction in general (it depends on the non-Archimedean structure of this set).

# 6. Conclusion

In the paper I have proposed a many-valued calculus in that illocutionary forces and performances are considered as modal operators of a special kind. As a result, I have constructed an easier formalization of illocutionary act theory than usual ones. This formalization could be applied in model-theoretic semantics of natural language and natural language programming.